\documentclass[aps,pra,twocolumn,showpacs,amsmath,a4paper,amssymb,superscriptaddress,10pt]{revtex4-1}
\usepackage{graphicx}
\usepackage{CJK}

\usepackage{amsmath,amssymb}
\usepackage{color}
\usepackage{soul}

\begin{document}
\begin{CJK*}{UTF8}{gbsn}

\title{Fragmentation of long-lived hydrocarbons after strong field ionization}

\author{Seyedreza Larimian}
\author{Sonia Erattupuzha}
\affiliation{Photonics Institute, Vienna University of Technology, A-1040 Vienna, Austria}

\author{Erik L\"otstedt}\email[Electronic address: ]{lotstedt@chem.s.u-tokyo.ac.jp}
\author{Tam\'as Szidarovszky}
\affiliation{Department of Chemistry, School of Science, The University of Tokyo, 7-3-1 Hongo, Bunkyo-ku, Tokyo 113-0033, Japan}

\author{Raffael Maurer}
\author{Stefan Roither}
\author{Markus Sch\"offler}
\author{Daniil Kartashov}
\author{Andrius Baltu\v{s}ka}
\affiliation{Photonics Institute, Vienna University of Technology, A-1040 Vienna, Austria}

\author{Kaoru Yamanouchi}
\affiliation{Department of Chemistry, School of Science, The University of Tokyo, 7-3-1 Hongo, Bunkyo-ku, Tokyo 113-0033, Japan}

\author{Markus Kitzler}\email[Electronic address: ]{markus.kitzler@tuwien.ac.at}
\affiliation{Photonics Institute, Vienna University of Technology, A-1040 Vienna, Austria}

\author{Xinhua Xie (谢新华)}\email[Electronic address: ]{xinhua.xie@tuwien.ac.at}
\affiliation{Photonics Institute, Vienna University of Technology, A-1040 Vienna, Austria}

\date{\today}

\begin{abstract}
We experimentally and theoretically investigated the deprotonation process on nanosecond to microsecond timescale in ethylene and acetylene molecules, following their double ionization by a strong femtosecond laser field. In our experiments we utilized coincidence detection with the reaction microscope technique, and found that both the lifetime of the long-lived ethylene dication leading to the delayed deprotonation and the relative channel strength of the delayed deprotonation compared to the prompt one have no evident dependence on the laser pulse duration and the laser peak intensity.
Quantum chemical simulations suggest that such delayed fragmentation originates from the tunneling of near-dissociation-threshold vibrational states through a dissociation barrier on a dication electronic state along C--H stretching.
Such vibrational states can be populated through strong field double ionization induced vibrational excitation on an electronically excited state in the case of ethylene, and through intersystem crossing from electronically excited states to the electronic ground state in the case of acetylene.
\end{abstract}

\maketitle
\end{CJK*}

\section{Introduction}

The timescale of chemical reactions depends on the properties of the involved atoms or molecules and the external reaction conditions.
In the past decades, femto-chemistry, based on the use of femtosecond laser pulses, has become a vivid research direction in probing and controlling chemical reactions  \cite{zewail1988laser,zewail2000femtochemistry}.
Traditionally, femto-chemistry employs temporally shaped femtosecond laser pulses to steer the nuclear vibrational dynamics to control the formation or breakage of molecular bonds \cite{assion1998control}.
In recent years, laser pulses with high intensity and ultrashort pulse duration were successfully applied to the control of molecular fragmentation and isomerization reactions through strong field ionization or excitation \cite{liu11CO,xie12prl2,kling13pccp,wells2013adaptive,alnaser2014subfemtosecond,xie14prx,xie14prl}.
Such molecular fragmentation is a process involving one or more bond breakages during or after the interaction of a molecule with a strong laser field \cite{yamanouchi02}.
Exposed to a strong laser field, a molecule can be multiply ionized through multiphoton, tunneling or over barrier ionization, resulting in a certain charged state, which may be dissociative, and lead to the breakup of the ion into several pieces.
The molecular fragmentation from a dissociative state, which is induced by strong field ionization, is in general an extremely fast process happening on femtosecond or picosecond timescales \cite{zyubina2005,Alnaser2006,Sandhu1081,osipov2008,Zhou2012}.
It has been reported that such fragmentation processes can be controlled by a strong laser field via enhancing or suppressing the population of corresponding electronic states \cite{xie12prl2,xie14prx,xie14prl}.
In the case of the strong field interaction with molecules, the contribution of electrons in low-lying molecular orbitals becomes important for the ionization process.
Removing electrons from low-lying molecular orbitals may put the molecular ion into a dissociative electronically excited state \cite{talebpour99,xie12prl2,xie14prl}.
The decay of such dissociative states in general happens directly after strong field ionization.
Due to the complexity of the potential energy structure, such states can cross with other electronic states, which may open up several possible fragmentation pathways \cite{gaire14pra}.

Reaction microscopy is one of the frequently applied techniques in the study of strong field induced molecular fragmentation \cite{doerner00,Ullrich03}.
Such molecular fragmentation processes happen on a much faster timescale than the nanosecond temporal resolution of the detector and data acquisition system \cite{yamanouchi02}.
However, in this paper by using reaction microscope technique we report the observation of fragmentation of long-lived hydrocarbon dications, which have lifetimes of hundreds of nanoseconds to microseconds, after ionization in a strong femtosecond non-resonant laser field.
Our results suggest that such fragmentation originates from high-lying vibrational states near the dissociation threshold which have a long lifetime of hundreds of nanoseconds up to a few microseconds.
We present quantum chemical simulations which indicate that such high-lying vibrational states could be populated through strong field double ionization induced vibrational excitation on an electronically excited state, or by intersystem crossing (transition between states of different spin multiplicity) \cite{Bixon1968} from an electronically excited state to the electronic ground state.

\section{Experiments}

In the experiments we employ a reaction microscope to measure the three-dimensional momentum vectors of ions from the laser-molecule interaction \cite{doerner00,Ullrich03}. We use few-cycle laser pulses with a pulse duration down to 4.5 fs (full width at half maximum of the peak intensity), which are generated by spectral broadening and recompression of 25 fs pulses from a Titanium-Sapphire laser amplifier system.
These pulses are guided into an ultrahigh vacuum reaction chamber ($\sim$1.3$\times$10$^{-10}$ mbar), and focused onto a cold supersonic jet of molecules (diameter of $\sim$170 $\mu$m).
The molecular jet is prepared by supersonic expansion of the gas from a nozzle with a diameter of 10 $\mu$m and collimated by a two-stage skimmer before being sent to the reaction chamber.
We carried out measurements on acetylene and ethylene.
Ions produced from the strong field interaction with the gas targets are guided to a time- and position-sensitive detector by a homogeneous electric field (23.1 V/cm).
The laser pulse duration is adjusted by introducing positive chirp via adding a certain amount of fused silica to the beam path of the shortest pulse. The laser peak intensity is varied by reflection off a glass block under different grazing incidence angles. Calibration of the peak intensity on target was done with an estimated precision of 10\% by separate measurements using single ionization of argon atoms in circularly polarized light \cite{Smeenk2011}. The duration and intensity stability of the pulses were monitored on a shot-to-shot basis by a stereo-above-threshold-ionization phase meter \cite{Sayler:11,Sayler:12}. More details on our experimental setup can be found in our previous publications \cite{xie14prx,Zhang2012,xie12prl1}.

\section{Results}

\subsection{Identification of the slow fragmentation channel}

\begin{figure}[htbp]
\centering
\includegraphics[width=0.48\textwidth,angle=0]{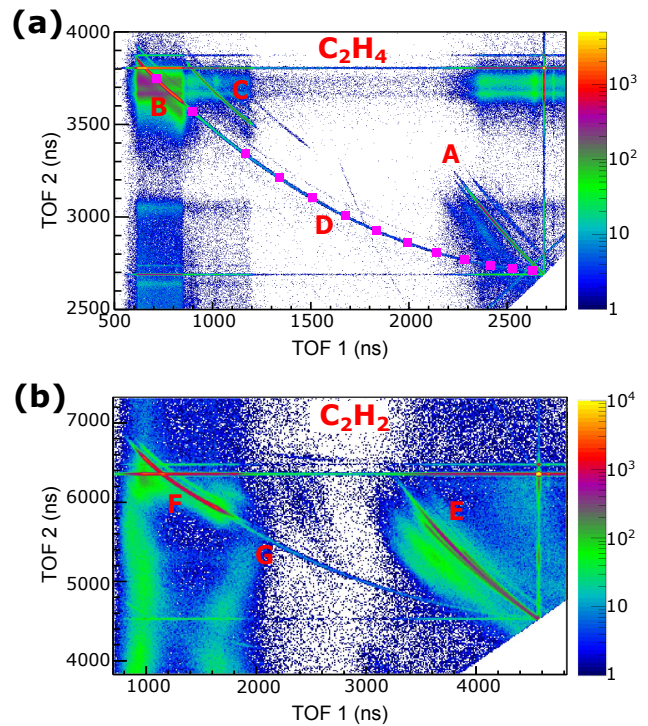}
\caption{(color online). Photoion-photoion-coincidence (PIPICO) distributions of ethylene (with the DC field strength of 23.1 V$/$cm) (a) and acetylene (with the DC field strength of 7.5 V$/$cm) (b) in a linearly polarized strong laser field with a peak laser intensity of 3$\times10^{14}$ W$/$cm$^2$ and a pulse duration of $4.5$ fs. A: CH$_2^+$+CH$_2^+$ B: C$_2$H$_3^+$+H$^+$ (prompt) C: C$_2$H$_2^+$+H$_2^+$  D: C$_2$H$_3^+$+H$^+$ (delayed) E: CH$^+$+CH$^+$ (C$^+$+CH$_2^+$) F: C$_2$H$^+$+H$^+$ (prompt) G: C$_2$H$^+$+H$^+$ (delayed). The magenta squares indicate the simulated PIPICO points for the delayed deprotonation process with a certain time delay between the ionization and the fragmentation process.}\label{fig:pipico}
\end{figure}

First, we focus on the experiments on ethylene molecules.
In previous experiments \cite{xie14prx}, we found that there are mainly three two-body fragmentation channels from ethylene dications after the strong field double ionization: deprotonation (C$_2$H$_3^+$+H$^+$), H$_2^+$ formation (C$_2$H$_2^+$+H$_2^+$), and symmetric break-up (CH$_2^+$+CH$_2^+$).
All three reactions proceed during or shortly after the laser interaction such that the fragmentation process takes place within the laser interaction region on a timescale much shorter than the resolution time of the detector and data acquisition system, which is about 1 nanosecond.
A typical measured photoion-photoion-coincidence (PIPICO) distribution of ethylene is presented in Fig.~\ref{fig:pipico}(a).
The PIPICO distribution presents the correlation between the time of flights (TOFs) of two detected particles on the detector.
Because of momentum conservation, the two-body fragmentation channels from ethylene dications appear as parabolic lines which are marked in Fig.~\ref{fig:pipico}(a).
We notice that the deprotonation channel has a weak but rather long tail while the PIPICO lines for the other two channels are short.
Such long tails have been already observed and studied in the dissociative double ionization of molecules with extreme-ultraviolet pulses \cite{alagia11jcp,alagia12jcp,alagia12cp}.
A long PIPICO line may indicate high kinetic energy release (KER) during the two-body fragmentation.
However, a long line corresponding to a high KER extends in both directions which is not the case in Fig.~\ref{fig:pipico}. In addition, we simulated the PIPICO line for a deprotonation process with a high KER and found that the measured long PIPICO line does not follow the simulated one. Therefore, we conclude that the long tail does not come from a fragmentation with high KER.

A long PIPICO line can be also formed by fragmentation happening long after the laser induced double ionization, during the flight of the dications to the detector.
To confirm such fragmentation process, we simulated the time of flight of C$_2$H$_3^+$ and H$^+$ for the case in which there is a time delay between ionization and fragmentation.
Before fragmentation, the ethylene dication flies towards the detector due to the weak DC field applied in the spectrometer.
At the time of fragmentation, the dication breaks into a proton and C$_2$H$_3^+$ which then both separately fly towards the detector and hit it at different times and positions according to their mass and charge.
We solve Newton's equations for the motion of these charged particles in a DC field and obtain their TOFs:
\begin{equation}
	\frac{1}{2}a_{z,i}(\text{TOF}_{i}-\text{T})^2 + v_{z,i}(\text{TOF}_{i}-\text{T}) + z = \text{L}
\end{equation}
where index $i=1,2$ refers to two different moieties after the fragmentation, $a_{z}=\frac{q.\text{E}}{m}$ ($q$: electric charge, $m$: particle mass, E: DC field strength) is the acceleration of the particle in the spectrometer direction by the applied DC electric field, T is the delay between the ionization and fragmentation (survival time of the dication), $v_{z}$ is the velocity of the particles at the time of fragmentation which can be found from initial conditions and the conservation of momentum at the time of fragmentation, $z=\frac{1}{2}a_{z,0}\text{T}^2$ is the position of the intact dication at the time of fragmentation ($a_{z,0}$ acceleration of the dication by the applied DC electric field) and L is the distance between the interaction point and the ion detector.
The initial momentum of the parent ion is assumed to be 0.
We vary the time delay and overlay the simulated TOFs of the two ions, which are shown as magenta points in Fig.~\ref{fig:pipico}(a).
These points perfectly overlap with the measured long PIPICO line.
This proves that the long PIPICO line originates from the deprotonation process of ethylene dications with a long survival time after double ionization.
In case of the measurements with acetylene, presented in Fig.~\ref{fig:pipico}(b), we observed a similar deprotonation process.
Additionally, such delayed deprotonation is observed also in measurements of ethylene and acetylene fragmentation induced by using circularly polarized laser pulses.
Summing up, we confirmed the observation of the delayed deprotonation process of ethylene and acetylene dications with a long survival time after strong field double ionization.

\subsection{Lifetime and KER distribution}

\label{subsec:Lifetime}
\begin{figure}[htbp]
\centering
\includegraphics[width=0.48\textwidth,angle=0]{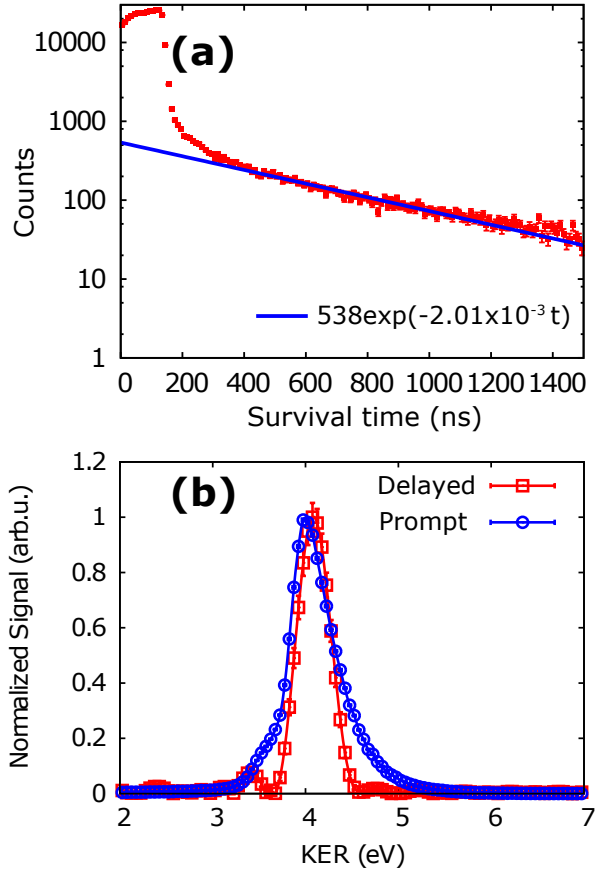}
\caption{(color online). (a) Retrieved survival time distribution of the ethylene dication which leads to the delayed deprotonation process with the exponential fitting of $538\exp(-2.01\times10^{-3}\text{t})$. (b) KER distributions of the prompt (blue circle line) and the delayed (red square line) deprotonation processes.}\label{fig:lifetime}
\end{figure}

From the measured data we select the long PIPICO line by applying the relation between the TOFs of the two detected ions derived from Newton's equations and the relation in $x$- and $y$-direction on the detector to ensure coincidence selection.
We retrieve the survival time of the ethylene dications using the relation derived from Newton's equations for the charged particles:
\begin{widetext}
\begin{equation}
	\text{T}=\frac{(a_{z,0}-a_{z,2})(a_{z,1}\text{TOF}_{1}^2-2\text{L}) - (a_{z,0}-a_{z,1})(a_{z,2}\text{TOF}_{2}^2-2\text{L})}{2(a_{z,0}-a_{z,2})(a_{z,0}-a_{z,1})(\text{TOF}_{2}-\text{TOF}_{1})}\label{equ:lifetime}
\end{equation}
\end{widetext}
The distribution of the survival time for the ethylene dication is shown in Fig.~\ref{fig:lifetime}(a).
Note that the peak of the signal for the survival time less than $200$ ns originates from the prompt fragmentation. This peak forms due to the application of Eq.~\ref{equ:lifetime} to the region where hits from the fast and slow process overlap.
However, to distinguish the desired signal we take into account only the data points with a survival time of longer than $400$ ns, which exclusively originate from the long-lived dications.
The figure clearly shows that the decay of the measured delayed deprotonation signal of ethylene dications is exponential.
We fit the measured signals with an exponential function $S(t)=S_0e^{-\alpha t}$, shown as the blue solid line in Fig.~\ref{fig:lifetime}(a), which yields the decay rate $\alpha=(2.01\pm0.05)\times10^{-3}$ ns$^{-1}$.
From the fitted parameter $\alpha$, we obtain the lifetime of ethylene dications for the delayed deprotonation channel as $1/\alpha=498\pm12$ ns.

The momentum of the proton in the detector plane, which is gained during the deprotonation process from long-lived dications, can be calculated from the measured survival time and the position on the detector.
In our measurement, the minimum data acquisition integration time is on the order of 1 ns which is much longer than the rotational period of the ethylene molecule which is about 18 ps \cite{rouzee06}. 
Therefore any laser induced alignment effect will be smearing out, which leads the 3D momentum distribution of the protons to be isotropic.
For a certain TOF, the retrieved momentum distribution in the detector plane is a projection of the 3D isotropic sphere onto a 2D plane.
Therefore, to get the correct 2D momentum distribution, an Abel-transform needs to be applied to the raw momentum distribution \cite{dribinski2002}.
From the retrieved momentum, we can then obtain the KER of the delayed deprotonation process which is shown in Fig.~\ref{fig:lifetime}(b).
The mean value of the KER is 4.1 eV (with a FWHM of 0.4 eV), which is almost the same as that of the prompt deprotonation process.
Additionally, by selecting a different survival time region we found that the KER has no dependence on the survival time of the ethylene dication.
This indicates that all slow deprotonation events originate from the same initial state.

\subsection{Dependence on laser parameters}

\begin{figure}[htbp]
\centering
\includegraphics[width=0.48\textwidth,angle=0]{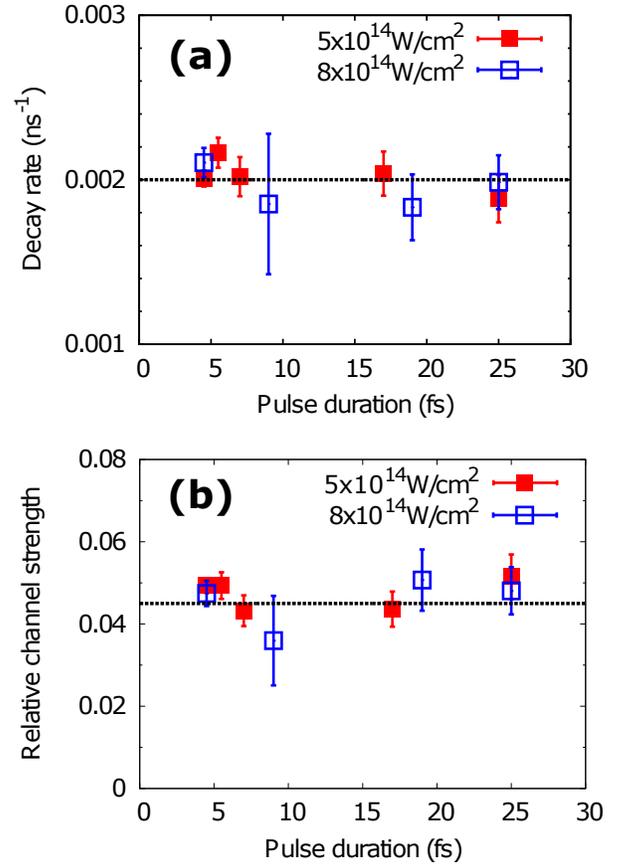}
\caption{(color online). (a) Retrieved decay rate of ethylene dication for the delayed deprotonation process as a function of the laser pulse duration for laser peak intensities of $5\times10^{14}$ W$/$cm$^2$ and $8\times10^{14}$ W$/$cm$^2$. (b) Retrieved relative channel strength of the delayed deprotonation process with respect to the prompt one.}\label{fig:laser}
\end{figure}

In previous studies we found that strong field induced molecular fragmentation processes strongly depend on laser parameters, such as pulse duration, peak intensity and carrier-envelope phase \cite{roither11prl,xie12prl2,xie14prx,xie14pra,xie15sr}.
As shown in Ref. \cite{xie14prx}, the relative channel strength between the deprotonation process and the stable ethylene dication has a strong dependence on the laser pulse duration.
To investigate the origin of the deprotonation process from long-lived dications, especially its relation to the prompt deprotonation process, we performed measurements on ethylene with pulse durations from 4.5 fs up to 25 fs with peak intensities of $5\times10^{14}$ W$/$cm$^2$ and $8\times10^{14}$ W$/$cm$^2$.
In the data analysis, we obtain the decay rate of the ethylene dication for the delayed deprotonation from the exponential fitting.
Because of the overlapping with the strong prompt deprotonation signal as shown in Fig.~\ref{fig:lifetime}(a), it is impossible to directly obtain the yield of the deprotonation channel from the long-lived dications.
However, we can get the yield by integrating the fitting function over survival time which leads to: $N_{0}=\int_0^\infty S_{0}\mathrm{e}^{-\alpha t}\,\mathrm{d}t=S_0/\alpha$.
For the prompt deprotonation channel, we performed the usual coincidence momentum retrieval with a time delay between ionization and fragmentation shorter than 1 ns and applied momentum conservation conditions in all three dimensions \cite{xie14prx}.
The relative channel strength of the delayed deprotonation to the prompt one is about 4.5\%.
The decay rate and the relative channel strength between them as a function of laser pulse duration are presented in Fig.~\ref{fig:laser}(a) and Fig.~\ref{fig:laser}(b), respectively, for two laser intensities.
It is shown that both the decay rate and the relative strength of these two processes have no evident dependence on the laser pulse duration and intensity.
No dependence of the decay rate on the laser parameters indicates that all signals of delayed deprotonation originate from the decay of the same vibronic state of the ethylene dication.
On the other hand, no dependence of the relative channel strength denotes that both processes originate from the same nuclear wave packet on the same potential energy surface, which is populated after the strong field double ionization.
No dependence on laser parameters also indicates that the vibronic states with a long lifetime, leading to the delayed deprotonation, are not populated by laser-induced excitation from lower-lying states.

\section{Discussions}

In the following we will discuss the mechanism behind the long lifetime of ethylene dications, leading to the delayed deprotonation process.

In recent studies we found that electronically excited states of the ethylene dication are dissociative and lead to fragmentation processes, such as deprotonation, symmetric breaking or H$_{2}^+$ formation \cite{xie12prl2,xie14prx}.
The electronic ground state of the ethylene dication is meta-stable and has a very long lifetime, which might be a candidate for the mentioned deprotonation process \cite{xie14prx}.
However, based on the same laser pulse duration and intensity dependence of the delayed and prompt deprotonation processes, we conclude that both processes may originate from
the same electronic state.
In Ref. \cite{xie14prx}, it was shown that the prompt deprotonation originates from electronically excited states populated by removal of at least one HOMO-1 electron, while the other excited states which involve removal of electrons from HOMO-2 preferentially lead to C--C breakage.
Therefore, the contribution from the electronic ground state can be ruled out.
On the other hand, for the electronically excited states which lead to the C--C breakage, there is no energy barrier along the C--C stretching coordinate, which supports the absence of a delayed C--C breakage process in our measurements.

\begin{figure}[ht]
\centering
\includegraphics[width=0.48\textwidth,angle=0]{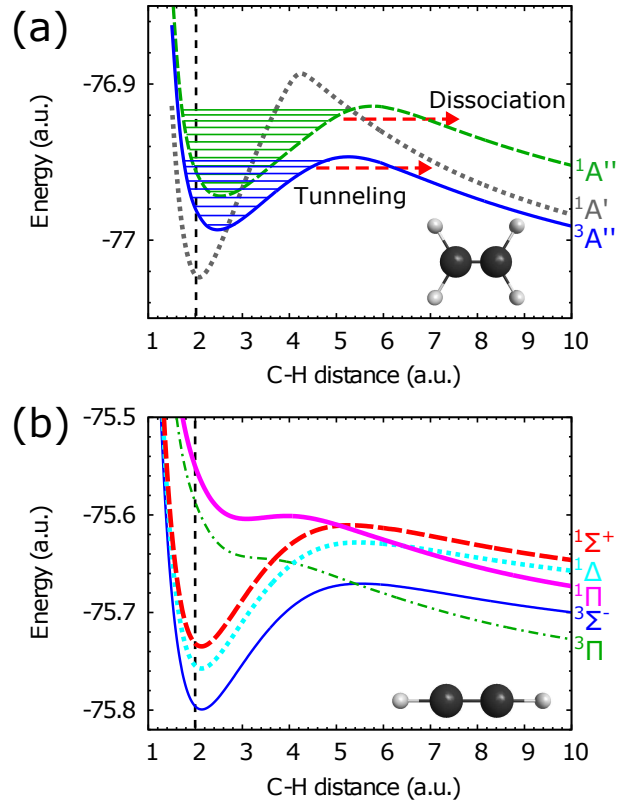}
\caption{(color online). (a) Calculated PECs along a C-H stretching coordinate of the planar ethylene dication. The gray dotted line corresponds to the electronic ground states, while the blue solid line and the green dashed line present the first and second electronically excited states, respectively. The horizontal lines represent for the vibrational levels. (b) Calculated PECs along a C-H streching coordinate of the acetylene dication, with all other coordinates in the neutral equilibrium geometry.
The vertical dashed lines are the equilibrium C--H distances of the neutral molecules.
\label{fig:pes}}
\end{figure}

To obtain insight into the electronically excited states which might lead to the long-lived dications we simulated the PECs of the ethylene dication with the planar geometry (the equilibrium geometry of neutral ethylene) along a C--H stretching coordinate.
Here the complete active space  self-consistent field method (6-311++G(d,p) basis set) as implemented in {\sc gamess} \cite{gamess} was used, with 10 active electrons in 12 orbitals.
A one-dimensional potential energy curve (PEC) was obtained by varying one of the C-H bond lengths in C$_2$H$_4^{2+}$,
while all other bond lengths and bond angles were kept fixed at the values corresponding to the equilibrium structure of neutral C$_2$H$_4$.
The PECs of the electronic ground state, the lowest electronically excited triplet state (formed by removing one HOMO and one HOMO-1 electron with the same spin) and the lowest electronically exited singlet state (formed by removing one HOMO and one HOMO-1 electron with opposite spin) are presented in Fig.~\ref{fig:pes}(a) as the gray line, the blue line and the green line, respectively.
The PECs of the two excited states show barriers along the C-H stretching coordinate with a height of about 1.5 eV and 1.2 eV, respectively, which might support long lifetime of dication prior to deprotonation.
When a nuclear wave packet is populated in a vibrational level close to the dissociation barrier, it can tunnel through the dissociation barrier leading to delayed deprotonation.

Utilizing the complex coordinate scaling method \cite{reinhardt1982complex}, we calculated the quasi-bound vibrational energies (relative to the minimum of the potential well) and lifetimes \cite{klaiman2012resonance} supported by the one-dimensional PECs for the first excited (triplet) state and the second excited (singlet) state in C$_2$H$_4^{2+}$, as presented in Tab.~\ref{tab:first} and Tab.~\ref{tab:second}.

\begin{table}[ht]
    \centering
    \caption{Vibrational levels along the C-H stretching coordinate on the first excited electronic state of the ethylene dication}
    \label{tab:first}
    \begin{ruledtabular}
    \begin{tabular}{ l | r | r | r }
    $\nu$ & Energy (eV) & Lifetime (ns) & FC factor \\ \hline
    1 & 0.078   & $>$10$^5$  & 7.70$\times 10^{-2}$  \\ \hline
    2 & 0.229   & $>$10$^5$  & 1.61$\times 10^{-1}$  \\ \hline
    3 & 0.379   & $>$10$^5$  & 1.92$\times 10^{-1}$  \\ \hline
    4 & 0.525   & $>$10$^5$  & 1.71$\times 10^{-1}$  \\ \hline
    5 & 0.668   & $>$10$^5$  & 1.27$\times 10^{-1}$  \\ \hline
    6 & 0.808   & $>$10$^5$  & 8.27$\times 10^{-2}$  \\ \hline
    7 & 0.942   & $>$10$^5$  & 4.94$\times 10^{-2}$  \\ \hline
    8 & 1.072   & $>$10$^5$  & 2.77$\times 10^{-2}$  \\ \hline
    9 & 1.194   & $>$10$^5$  & 1.50$\times 10^{-2}$  \\ \hline
    10 & 1.309   & 1.30$\times 10^{3}$  & 7.92$\times 10^{-3}$  \\ \hline
    11 & 1.413   & 3.00$\times 10^{-1}$  & 4.18$\times 10^{-3}$ \\ \hline
    12 & 1.501   & 2.17$\times 10^{-4}$  & 4.18$\times 10^{-3}$  \\
    \end{tabular}
    \end{ruledtabular}
\end{table}

\begin{table}[ht]
    \centering
    \caption{Vibrational levels along the C-H stretching coordinate on the second excited electronic state of the ethylene dication}
    \label{tab:second}
    \begin{ruledtabular}
    \begin{tabular}{  l | r | r | r }
    $\nu$ & Energy (eV) & Lifetime (ns) & FC factor \\ \hline
    1 & 0.084   & $>$10$^5$  & 1.25$\times 10^{-1}$  \\ \hline
    2 & 0.248   & $>$10$^5$  & 2.17$\times 10^{-1}$  \\ \hline
    3 & 0.407   & $>$10$^5$  & 2.17$\times 10^{-1}$  \\ \hline
    4 & 0.561   & $>$10$^5$  & 1.63$\times 10^{-1}$  \\ \hline
    5 & 0.709   & $>$10$^5$  & 1.04$\times 10^{-1}$  \\ \hline
    6 & 0.849   & $>$10$^5$  & 5.91$\times 10^{-2}$  \\ \hline
    7 & 0.980   & 3.46$\times 10^{4}$  & 3.61$\times 10^{-2}$  \\ \hline
    8 & 1.099   & 2.53  & 1.63$\times 10^{-2}$  \\ \hline
    9 & 1.201   & 1.19$\times 10^{-3}$  & 8.15$\times 10^{-3}$  \\
    \end{tabular}
    \end{ruledtabular}
\end{table}

We obtain lifetimes of 1.3 $\mu$s and 2.53 ns for the third and the second highest vibrational levels of the first and the second excited states, respectively, which are on the same timescale of the experimentally obtained value of 498 ns.
The theoretical calculations thus support the hypothesis of near-dissociation-threshold tunneling as the origin of the long-lived dications.
By slightly varying the PECs, we confirmed  that the lifetimes of the high-lying vibrational states are very sensitive to the exact shape of the potential.
A slight change of the barrier height can result in an order-of-magnitude change of the lifetime. Therefore, it is not meaningful to make a quantitative comparison of the computed lifetimes with the experimentally determined values.
On the other hand, the lifetimes of the highest vibrational levels of the two excited states are on the order of femtosecond and picoseconds, which supports the prompt deprotonation process.
Additionally, the prompt deprotonation could also originate from nuclear wave packets which are populated into the continuum above the dissociation barrier after double ionization.

From the calculated PECs, we estimate the KERs for the fragmentation from the highest vibrational levels of the two excited states to the dissociation limit.
It yields 3.7 eV and 3.5 eV for the delayed fragmentation from the first and the second electronic excited states, which is slightly smaller than the measured value of 4.1 eV.

\subsection{How are high-lying vibrational states populated?}
\label{Subsec:popmechanism}

Now we come to the question how such vibrational states can be populated during the interaction of a strong laser field with an ethylene molecule.
Previous studies indicate that electronically excited states of an ethylene dication can be populated directly through strong field double ionization by removing one or two electrons from lower lying molecular orbitals \cite{xie12prl2,xie14prx,xie14prl}.

In our measurements, the double ionization process happens within a few-cycle laser pulse, during which the nuclei can be considered frozen in the neutral equilibrium geometry.
From the simulations, as listed in Tab.~\ref{tab:first} and Tab.~\ref{tab:second}, we found that the Franck-Condon (FC) factors between the vibronic ground state of the neutral ethylene and vibrational excited states of the ethylene dication on the electronically excited states are on the order of 10$^{-2}$ for the high-lying vibrational states that can dissociate through tunneling.
This indicates that the high-lying vibrational states can be directly populated through strong field double ionization through removing at least one HOMO-1 electron.

\subsection{Comparison between ethylene and acetylene}

From the acetylene measurements we retrieved the decay rate of long-lived acetylene dication for the deprotonation process. We found it as $5.9(\pm0.5)\times10^{-4}$ ns$^{-1}$ corresponding to a mean survival time of 1695$\pm$144 ns, which is much longer than that of the ethylene dication.
The relative channel strength of the delayed deprotonation to the prompt one for acetylene is 2.9\%, which is almost the same as that of ethylene.

Until now, we presented the observation of similar deprotonation processes in ethylene and acetylene.
In the neutral state, the main difference between the two species is that ethylene has a C--C double bond and acetylene has a C--C triple bond.
Their dications also have significant difference, due to the different electronic configuration.

As presented in Fig.~\ref{fig:pes}(b), the three lowest electronic states of the acetylene dication are meta-stable.
The lowest state is the electronic ground dication state which is a triplet state.
It is formed by removal of two electrons with the same spin, one from HOMO and the other one from the degenerate HOMO-1.
The other two states are, however, formed by removal of two electrons with different spins either both from HOMO or one from HOMO and the other from HOMO-1.
All three electronic states can be populated via strong laser field induced double ionization.
Similar to the electronic ground state of the ethylene dication, the low-lying vibrational levels will be dominantly populated during the double ionization process from the neutral acetylene due to rather small FC factors between high-lying vibrational states of these three states and the vibrational ground state of the neutral acetylene.
The population in the low-lying vibrational states of the electronic ground state contributes to the measured stable acetylene dications whose survival time are longer than their TOF to the detector.
Similar to ethylene, to support the delayed deprotonation process, the nuclear wave packet needs to be prepared on a high-lying vibrational state near the dissociation threshold of an electronic state in acetylene dication.
We note that direct vibrational excitation through ionization to near-threshold vibrational states on an electronic state of the acetylene dication is very unlikely due to the small FC factors.
However, the population in the two upper meta-stable electronic states can possibly decay nonradiatively into the high-lying vibrational states of the electronic ground state via a spin flip transition (intersystem crossing).
Some of the populated high-lying vibrational states can be close to the dissociation threshold along the C--H stretching and the nuclear wave packet might tunnel through the dissociation barrier and lead to slow deprotonation.

Finally, it is noted, that an alternative explanation that may explain the slow deprotonation process both for ethylene and acetylene is over-the-barrier dissociation.
In this case, the dissociation in principle might proceed on a metastable electronic state, from a vibrational state that has an energy slightly larger than the potential barrier.
Although the one-dimensional model for the deprotonation along the C-H stretching coordinate suggests that the FC factor for such above barrier vibrational states should be very small for PECs similar to the neutral ground state PEC (such as the electronic ground state of ethylene and the three lowest lying electronic states of acetylene), that is not necessarily the case.
If the possibility of excitation in other vibrational modes is also considered, the multidimensional vibrational wave packet could have larger energy than the barrier along the C-H stretching, while simultaneously its projection onto the C-H stretching coordinate could have a simple structure corresponding to a significant FC factor.
This way, both above barrier continuum states leading to fast deprotonation, and above barrier Feshbach type resonances \cite{moiseyev98} leading to slow deprotonation could be formed. Quantitative simulation of this mechanism, however, requires accurate global potential energy surfaces and multidimensional wave packet propagations, which is a formidable task, and was not attempted in this work.
Using the Rice-Ramsperger-Kassel-Marcus (RRKM) theory, in \cite{zyubina2005}, the rate constant $k$ for the dissociation of C$_2$H$_2^{2+}$$\to$C$_2$H$^{+}$$+$H$^+$ on the electronic ground triplet PES was estimated to be $k=2.2\times 10^{12}$ s$^{-1}$ for a total internal energy of about 1 eV above the dissociation barrier.
Although this value of $k$ is too large to explain the experimentally measured $\mu$s-order lifetime, we note that $k$ is expected to decrease if the total energy decreases to a value only slightly larger than the barrier height.

\section{Conclusion}

In conclusion, we experimentally and theoretically studied the slow fragmentation of hydrocarbon dications following double ionization by a strong laser field. A possible scenario of the slow deprotonation process in ethylene, which is consistent with the simple theoretical model applied in this work is as follows.
In the first step, an ethylene molecule is doubly ionized by removing at least one HOMO-1 electron. 
This prepares the dication in an electronically excited state.
Due to the Frank-Condon principle, high-lying vibrational states on the electronically excited states will be populated through strong field double ionization.
Such high-lying vibrational states can later tunnel through the dissociation barrier along the C--H stretching coordinate which leads to the slow deprotonation process.
In acetylene, the slow deprotonation process probably occurs from the near-dissociation-threshold vibrational states in the electronic ground state.
The mechanism for populating such states can be explained by vibrational excitation through intersystem transitions from electronically excited states to the electronic ground state.
Since vibrational excitation through double ionization and intersystem crossing are common processes in a variety of molecular systems, the slow deprotonation process observed in our experiment should play an important role in the strong field interaction with a variety of polyatomic molecules.
Such a process can potentially be used in the coherent control of chemical reactions, and this work may trigger further experimental and theoretical studies on strong field induced molecular reactions.

We thank Dr. Artem Rudenko for helpful discussions. This work was financed by the Austrian Science Fund (FWF) under grants P25615-N27, P28475-N27, P21463-N22, P27491-N27 and SFB-F49 NEXTlite, by a starting grant from the ERC (project CyFi), by the Ministry of Education, Culture, Sports, Science and Technology (MEXT), Japan (Grant-in-Aid for Specially Promoted Research No. 19002006), and by the Grant-in-Aid (Tokubetsu Kenkyuin Shorei-hi) scientific research fund of JSPS (Japan Society for the Promotion of Science).

\end{document}